\documentclass[12pt]{article}
\usepackage{amssymb}
\usepackage{graphics}
\usepackage{epsfig}

\DeclareFontFamily{U}{rsf}{}
\DeclareFontShape{U}{rsf}{m}{n}{
  <5> <6> rsfs5 <7> <8> <9> rsfs7 <10-> rsfs10}{}
\DeclareMathAlphabet\Scr{U}{rsf}{m}{n}

\catcode`\@=11
\def\@citex[#1]#2{%
\if@filesw \immediate \write \@auxout {\string \citation {#2}}\fi
\@tempcntb\m@ne \let\@h@ld\relax \def\@citea{}%
\@cite{%
  \@for \@citeb:=#2\do {%
    \@ifundefined {b@\@citeb}%
      {\@h@ld\@citea\@tempcntb\m@ne{\bf ?}%
      \@warning {Citation `\@citeb ' on page \thepage \space undefined}}%
      {\@tempcnta\@tempcntb \advance\@tempcnta\@ne%
      \@tempcntb\number\csname b@\@citeb \endcsname \relax%
      \ifnum\@tempcnta=\@tempcntb 
        \ifx\@h@ld\relax%
          \edef \@h@ld{\@citea\csname b@\@citeb\endcsname}%
        \else%
          \edef\@h@ld{\ifmmode{-}\else--\fi\csname b@\@citeb\endcsname}%
        \fi%
      \else
        \@h@ld\@citea\csname b@\@citeb \endcsname%
        \let\@h@ld\relax%
      \fi}%
    \def\@citea{,\penalty\@highpenalty\,}%
  }\@h@ld
}{#1}}

\def\@citeb#1#2{{[#1]\if@tempswa , #2\fi}}
\def\@citeu#1#2{{$^{#1}$\if@tempswa , #2\fi }}
\def\@citep#1#2{{#1\if@tempswa , #2\fi}}

\def\bcites{         
        \catcode`\@=11
        \let\@cite=\@citeb
        \catcode`\@=12
}

\def\upcites{         
        \catcode`\@=11
        \let\@cite=\@citeu
        \catcode`\@=12
}

\def\plaincites{      
        \catcode`\@=11
        \let\@cite=\@citep
        \catcode`\@=12
}

\newcount\hour
\newcount\minute
\newtoks\amorpm
\hour=\time\divide\hour by 60
\minute=\time{\multiply\hour by 60 \global\advance\minute by-\hour}
\edef\standardtime{{\ifnum\hour<12 \global\amorpm={am}%
        \else\global\amorpm={pm}\advance\hour by-12 \fi
        \ifnum\hour=0 \hour=12 \fi
        \number\hour:\ifnum\minute<10 0\fi\number\minute\the\amorpm}}
\edef\militarytime{\number\hour:\ifnum\minute<10 0\fi\number\minute}

\def\draftlabel#1{{\@bsphack\if@filesw {\let\thepage\relax
   \xdef\@gtempa{\write\@auxout{\string
      \newlabel{#1}{{\@currentlabel}{\thepage}}}}}\@gtempa
   \if@nobreak \ifvmode\nobreak\fi\fi\fi\@esphack}
        \gdef\@eqnlabel{#1}}
\def\@eqnlabel{}
\def\@vacuum{}
\def\marginnote#1{}
\def\draftmarginnote#1{\marginpar{\raggedright\scriptsize\tt#1}}
\def\draft{
        \pagestyle{plain}
        \overfullrule=2pt
        \oddsidemargin -.5truein
        \def\@oddhead{\sl \phantom{\today\quad\militarytime} \hfil
        \smash{\Large\sl DRAFT} \hfil \today\quad\militarytime}
        \let\@evenhead\@oddhead
        \let\label=\draftlabel
        \let\marginnote=\draftmarginnote
        \def\ps@empty{\let\@mkboth\@gobbletwo
        \def\@oddfoot{\hfil \smash{\Large\sl DRAFT} \hfil}
        \let\@evenfoot\@oddhead}
        \def\@eqnnum{(\theequation)\rlap{\kern\marginparsep\tt\@eqnlabel}%
        \global\let\@eqnlabel\@vacuum}  }

\def\section{\@startsection {section}{1}{\z@}{3.ex plus 1ex minus
 .2ex}{2.ex plus .2ex}{\large\bf}}
\def\subsection{\@startsection{subsection}{2}{\z@}{2.75ex plus 1ex minus
 .2ex}{1.5ex plus .2ex}{\bf}}

\def\appendix{{\newpage\section*{Appendix}}\let\appendix\section%
        {\setcounter{section}{0}
        \gdef\thesection{\Alph{section}}}\section}

\def\abstract{\if@twocolumn
\section*{Abstract}
\else 
\begin{center}
{\bf Abstract\vspace{-.5em}\vspace{0pt}}
\end{center}
\quotation
\fi}

\catcode`\@=12

\newcommand{\beq}{\begin{equation}}
\newcommand{\eeq}{\end{equation}}
\newcommand{\beqa}{\begin{eqnarray}}
\newcommand{\eeqa}{\end{eqnarray}}

\newcommand{\R}{{\bf R}}

\newcommand{\be}{\begin{eqnarray}}
\newcommand{\ee}{\end{eqnarray}}
\newcommand{\nn}{\nonumber}

\def\lae{\mathrel{\mathop{\smash{\lower .5 ex \hbox{$\stackrel<\sim$}}}}}
\def\lae{\mathrel{\mathop{\smash{\lower .5 ex \hbox{$\stackrel>\sim$}}}}}

\def\Tr{{\rm Tr}}

\def\l:{\mathopen{:}\,}
\def\r:{\,\mathclose{:}}

\catcode`\@=12

\bcites

\catcode`\@=11
\def\theequation{\arabic{equation}}
\@addtoreset{equation}{section}
\@addtoreset{footnote}{section}
\@addtoreset{footnote}{subsection}
\catcode`\@=12

\newcommand{\ft}[2]{{\textstyle\frac{#1}{#2}}}
\newcommand{\eqn}[1]{(\ref{#1})}
\def\Dslash{\,\,{\raise.15ex\hbox{/}\mkern-12mu D}}
\def\Dbarslash{\,\,{\raise.15ex\hbox{/}\mkern-12mu {\bar D}}}
\def\delslash{\,\,{\raise.15ex\hbox{/}\mkern-9mu \partial}}
\def\delbarslash{\,\,{\raise.15ex\hbox{/}\mkern-9mu {\bar\partial}}}
\def\pslash{\,\,{\raise.15ex\hbox{/}\mkern-9mu p}}
\def\calDslash{\,\,{\raise.15ex\hbox{/}\mkern-12mu {\cal D}}}

\begin{document}
\pagestyle{plain}
\setcounter{page}{1}
\newcounter{bean}
\baselineskip16pt

\begin{titlepage}

\begin{center}

\hfill\today\\
\hfill  TIT/HEP-545; DAMTP-2005-75\\

\vskip 3.0 cm {\Large Skyrmions from Instantons inside Domain
Walls} \vskip 1 cm
{Minoru Eto${}^1$, Muneto Nitta${}^1$, Keisuke Ohashi${}^1$ and David Tong${}^2$}\\
\vskip 1cm {\sl ${}^1$ Department of Physics, Tokyo Institute of Technology, \\
Tokyo, 152-8551, Japan.\\}
{\tt meto,nitta,keisuke@th.phys.titech.ac.jp}\\
\vskip .5cm
{\sl ${}^2$ Department of Applied Mathematics and Theoretical Physics, \\
University of Cambridge, CB3 0WA, UK.} \\
{\tt d.tong@damtp.cam.ac.uk}

\end{center}

\vskip 1.5 cm
\begin{abstract}
Some years ago, Atiyah and Manton described a method to construct
approximate Skyrmion solutions from Yang-Mills instantons. Here we
present a dynamical realization of this construction using domain
walls in a five-dimensional gauge theory. The non-abelian gauge
symmetry is broken in each vacuum but restored in the core of the
domain wall, allowing instantons to nestle inside the wall. We
show that the worldvolume  dynamics of the wall is given by the
Skyrme model, including the four-derivative term, and the
instantons appear as domain wall Skyrmions.
\end{abstract}
\end{titlepage}

\setcounter{section}{1}
\subsection*{\it Introduction}

Consider an $SU(2)$ gauge potential $A_\mu$ over Euclidian $\R^4$.
The path ordered holonomy along, say, the $x^4$ axis defines a
group valued field $g({\bf x})$ over $\R^3$,
\be g({\bf x})={\cal
P}\,\exp\left(i\int_{-\infty}^{+\infty}dx^4\,A_4({\bf
x},x^4)\right)\ .\label{am}\ee
We require that $A_{\mu}$ decays suitably rapidly as ${\bf
x}\rightarrow \infty$ so that $g({\bf x})\rightarrow 1$. Then,
regarding $g({\bf x})$ as the Skyrme field, Yang-Mills
configurations in $\R^4$ with instanton number $I$ map to Skyrme
configurations in $\R^3$ with baryon number $B=I$.

Equation \eqn{am} does not map solutions of the Yang-Mills
equations to solutions of the Skyrme equations. Indeed, while the
explicit form of the instanton is well known, no analytic solution
exists for the Skyrmion. Nevertheless, Atiyah and Manton showed
that the Yang-Mills instanton gives a remarkably good
approximation to the Skyrmion \cite{am}. More precisely, the
moduli space of $SU(2)$ instantons of charge $I$ has dimension
$8I$. After losing the translational mode along $x^4$, the map
\eqn{am} gives an $8I-1$ dimensional space of Skyrme
configurations. Minimizing the Skyrme energy functional over this
space results in an approximate solution which, in the case of a
single $B=I=1$ Skyrmion, has energy about $1\%$ above the true
solution. This construction has been studied for higher $B$ and
extended in other directions in \cite{1,2,3,theo,4,5}

The purpose of this paper is to provide a dynamical realization of
this correspondence, with instantons appearing as ``domain wall Skyrmions''. 
We construct a domain wall in a five dimensional Yang-Mills-Higgs theory whose
low-energy dynamics is described by the Skyrme model, complete
with four-derivative term. We show that the Skyrmions on the
domain wall worldvolume are instanton particles in five
dimensions, taking refuge inside the domain wall from the broken
gauge symmetry of the bulk.

There has been some interest of late in realizing skyrmions as 
instantons in a higher dimensional space, including models 
using orbifold deconstruction \cite{hill,son} and the AdS/CFT 
correspondence \cite{sugimoto}. 

\subsection*{\it A Five Dimensional Theory}

Our starting point is a five dimensional $U(2)$ gauge theory with
a single real scalar field $\phi$ transforming in the Lie algebra valued 
adjoint representation of the gauge group, and four complex scalar fields
$q_i$ $i=1,\dots,4$, each transforming in the fundamental
representation. The Lagrangian for our system is
\be {\cal L}=\frac{1}{2e^2}\Tr(-\ft12F_{\mu\nu}F^{\mu\nu}+({\cal
D}_\mu\phi)^2)+\sum_{i=1}^4(|{\cal D}_\mu q_i|^2
-q_i^\dagger(\phi-m_i)^2q_i)
-\frac{e^2}{2}\Tr(\,\sum_{i=1}^4q_i\otimes q_i^\dagger -v^2)^2
\nn\ee
where $\mu=0,1,\ldots, 4$ as befits $d=4+1$ Minkowski space. Both
the masses $m_i$ and the Higgs vacuum expectation value $v^2$ are
real parameters that are understood to come with an implicit
$2\times 2$ unit matrix. (e.g $v^2=v^2{\bf 1}_2$). If the masses
$m_i$ are set to zero, then this Lagrangian enjoys an $SU(4)_F$
flavor symmetry, rotating the $q_i$. For generic $m_i$ this is
broken to $U(1)_F^3$. In this paper we will be interested in an
intermediate situation in which $m_1=m_2=m$, and $m_3=m_4=-m$. The
surviving flavor symmetry in this case is $S[U(2)_{F_1}\times
U(2)_{F_2}]$. For notational purposes it will prove useful to 
separate these fundamental fields into two pairs: $Q_1=(q_1,q_2)$
and $Q_2=(q_3,q_4)$. We will write each of these as $2\times 2$
matrices, $(Q_1)^a_{\ i}$ and $(Q_2)^a_{\ i}$, with $a=1,2$ the
gauge index and $i=1,2$ denoting flavor. Then the action of the
$U(2)_G\times S[U(2)_{F_1}\times U(2)_{F_2}]$ gauge and flavor
symmetries on the scalars is given by
\be \phi\rightarrow U\phi U^\dagger\ \ \ \ , \ \ \ \
Q_1\rightarrow UQ_1V_1^\dagger\ \ \ \ \ Q_2\rightarrow
UQ_2V_2^\dagger\label{trans}\ee
where $U\in U(2)_G$ is a gauge symmetry and $V_i\in U(2)_{F_i}$
are flavor symmetries.

Domain walls in this theory with degenerate masses were studied
previously by Shifman and Yung \cite{sy} and further in
\cite{nitta}. We will be interested in walls interpolating between
two specific vacua of the theory, each of which is isolated with a
mass gap. They are:
\begin{itemize}
\item Vacuum 1: $\phi=m{\bf 1}_2$,\  $Q_1=v{\bf 1}_2$ and $Q_2=0$.
In this vacuum the  $U(2)_G\times SU(2)_{F_1}$ is spontaneously
broken to the diagonal $SU(2)_{\rm diag}$, while the $U(2)_{F_2}$
flavor symmetry survives.

\item Vacuum 2: $\phi=-m{\bf 1}_2$,\ $Q_1=0$ and $Q_2=v{\bf 1}_2$.
In this vacuum the  $U(2)_G\times SU(2)_{F_2}$ is spontaneously
broken to the diagonal $SU(2)_{\rm diag}$, while the $U(2)_{F_1}$
flavor symmetry survives.
\end{itemize}
Notice that each of these vacua lies in a "color-flavor" locked
phase, where a flavor rotation requires a matching color rotation.
However, a different flavor symmetry is locked with color in each
vacuum. This will prove important in the following.

There is also  a third branch of vacua in which $Q_1$ and
$Q_2$ are both non-zero and there is no mass gap. Although these
vacua play an interesting role in the dynamics of domain walls
\cite{sy}, they will not be of immediate interest for our story
and will enter our discussion only in passing.

\subsection*{\it The Domain Wall}

\begin{figure}[htb]
\begin{center}
\epsfxsize=3.2in\leavevmode\epsfbox{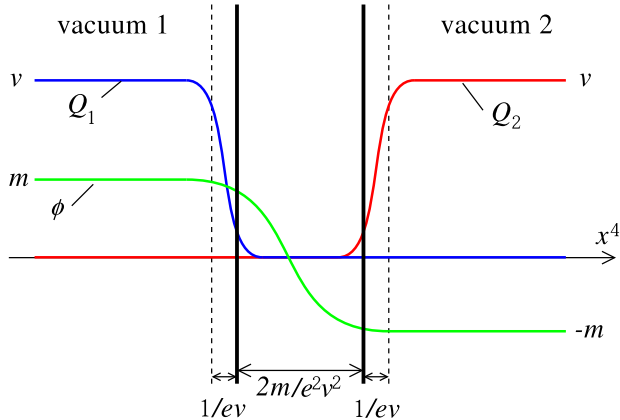}
\end{center}
\caption{The three layer structure of the domain wall when
$e^2v^2\ll m^2$.}
\end{figure}

Various parameters in our Lagrangian have been finely tuned so
that the theory admits a completion to one with eight
supercharges. However, supersymmetry will play no crucial role in
this paper and all arguments below depend only on the flavor and
gauge  symmetries of the model. Nevertheless, we stick with the
supersymmetric theory since domain walls are somewhat simpler in
this context.

We will examine the properties of the domain wall interpolating
between the two vacua described above in, say, the $x^4$
direction. The worldvolume of these domain walls is ${\bf
R}^{1,3}$ Minkowski space, spanned by $x^\alpha$ with
$\alpha=0,1,2,3$ and to find static configurations we set
$\partial_\alpha=A_\alpha=0$. One of the advantages of working
with the supersymmetric theory is that the second order equations
of motion can be integrated once to yield first order Bogomolnyi
equations,
\be {\cal D}_4\phi=e^2(Q_1Q_1^\dagger + Q_2Q_2^\dagger-v^2{\bf
1}_2)\ \ , \ \  {\cal D}_4Q_1=(\phi-m)Q_1\ \ ,\ \  {\cal
D}_4Q_2=(\phi+m)Q_2\label{dw}\ee
with boundary conditions imposed so that the fields asymptote to
Vacuum 1 as $x^4\rightarrow -\infty$ and Vacuum 2 as
$x^4\rightarrow +\infty$. Solutions to these equations are domain
walls with tension given by $T_{\rm wall}=4v^2m$.

The equations \eqn{dw} have an interesting space of solutions but
there are a particularly simple subset of solutions that will
interest us here \cite{sy}. We can satisfy the $U(2)$ group
structure of these equations by setting all fields proportional to
the $2\times 2$ unit matrix,
\be \phi=\phi(x^4)\,{\bf 1}_2\ \ ,\ \ Q_1=Q_1(x^4)\,{\bf 1}_2\ \
,\ \  Q_2=Q_2(x^4)\,{\bf 1}_2\ \ {\rm and}\ \ A_4=0\label{ans}\ee
in which case the equations \eqn{dw} reduce to those of an abelian
gauge theory studied previously in \cite{me,eto,sy2,sakai}. The
spatial profile of this domain wall is rather interesting and
depends on the dimensionless\footnote{Recall that in $d=4+1$, the
engineering dimensions of the various quantities are $[e^2]=-1$,
$[v^2]=3$ and $[m]=1$.} ratio $m^2/e^2v^2$. When $e^2v^2\ll m^2$,
the wall exhibits a three layer structure shown in the figure. The
fundamental fields $Q_1$ and $Q_2$ drop to zero over a thin shell
of width $1/ev$. The scalar $\phi$ takes a more leisurely path,
interpolating between its expectation values $\pm m$ over a length
$L_{\rm wall}\sim 2m/e^2v^2$. Note that in the middle of the
domain wall, the $U(2)$ gauge symmetry is naively restored. Of
course, this isn't true over distances larger than $L_{\rm wall}$ for which
long range non-abelian fields are affected by the boundary
conditions imposed by the bulk. In contrast, in the other limit
$e^2v^2\gg m^2$, the three layer structure is lost and all fields
vary over a length $1/m$.

Having found one solution using the ansatz \eqn{ans}, we may 
now act with the $U(2)_G\times S[U(2)_{F_1}\times U(2)_{F_2}]$ symmetry 
action to sweep out further solutions. 
The transformation of the scalar fields is
given in \eqn{trans} while the requirement that the vacua at left
and right infinity are untouched translates into the boundary
condition on the gauge transformation $U\in U(2)_G$,
\be U(x^4)\rightarrow\left\{\begin{array}{cr} V_1 & \ \ \ \ \
x^4\rightarrow -\infty \\ V_2 & x^4\rightarrow
+\infty\end{array}\right.\ \ . \label{lock}\ee
Under such a gauge transformation, the $A_4$ component of the
gauge field also picks up a contribution,
\be A_4(x^4)=-i (\partial_4 U)U^\dagger\ .\ee
Because the gauge symmetry is locked with a different flavor
symmetry at left and right infinity, a fact reflected in the
boundary condition \eqn{lock}, the holonomy of $A_4$ along the
$x^4$ direction transverse to the wall is non-vanishing. We have,
\be g={\cal
P}\exp\left(i\int_{-\infty}^{+\infty}dx^4\,A_4(x^4)\right)=V_2\,V_1^{-1}\
.\label{notam}\ee
%
This $U(2)$ valued field $g$ plays the role of the gauge invariant
coordinate \cite{sy} on the moduli space ${\cal M}$ of the domain
wall,
\be {\cal M}\cong\frac{U(2)_{F_1}\times U(2)_{F_2}}{U(2)_G}\cong
U(2)\ . \ee
The solution of the form \eqn{ans} corresponds to the point $g=1$.
All other solutions on ${\cal M}$ are related by a symmetry
transformation.

The domain wall has further moduli. There is, of course, the
overall center of mass of the wall. There are also zero modes
corresponding to the domain wall splitting into two walls, each
with tension $\ft12 T_{\rm wall}$, and with the third branch of
vacua mentioned above lying between them. We may consistently
ignore these moduli for the purpose of our story. In particular,
the moduli corresponding to splitting into two walls would not
exist in a non-supersymmetric theory.

Note the similarity between the coordinate on the moduli space of
domain walls \eqn{notam} and the Atiyah-Manton equation \eqn{am}.
The difference is that the collective coordinate $g$ of the domain
wall defined in \eqn{notam} does not yet depend on the wall
worldvolume coordinates $x^\alpha$ with $\alpha=0,1,2,3$. We shall
now rectify this.

\subsection*{\it The Skyrme Model on the Domain Wall}

In the spirit of the moduli space approximation, we promote the
collective coordinates $g$ to fields on the domain wall
worldvolume: $g\rightarrow g(x^\alpha)$, with $\alpha=0,1,2,3$.
The low-energy dynamics on the domain wall is derived by inserting
this varying collective coordinate back into the action, together
with a suitable ansatz for the gauge fields $A_\alpha$, parallel
to the domain wall, which are sourced by $\partial_\alpha g$. We now 
describe this procedure.

Given the symmetries of our model, the low-energy dynamics of the
collective coordinate $g$ at the two-derivative level is fixed to
be the $U(2)$ chiral Lagrangian. We need only determine the
overall coefficient "$f_\pi^2/16$". This can be achieved by
studying the fluctuations about the simplest solution \eqn{ans},
corresponding to the point $g=1$ in moduli space. The zero modes
may be written as
\be \delta\phi=i[\Gamma,\phi]\ \ ,\ \ \delta Q_i=i\Gamma
Q_i-iQ_i\Omega_i\ \ \  {\rm and}\  \ \delta A_4={\cal
D}_4\Gamma\label{fluc}\ee
where $U(x^4)=e^{i\Gamma(x^4)}$ is the gauge transformation and
$V_i=e^{i\Omega_i}$ are the two flavor transformations.  The gauge
transformation is restricted to obey the boundary conditions
\eqn{lock} which, in infinitesimal form, read,
\be \Gamma(x^4)\rightarrow\left\{\begin{array}{ll} \Omega_1 \ \ \
\ \ & x^4\rightarrow -\infty \\ \Omega_2 & x^4\rightarrow +\infty
\end{array}\right.\ \ .\ee
The full $x^4$ dependence of the gauge transformation
$\Gamma(x^4)$ is determined by a suitable background gauge fixing
condition which, for the fluctuations \eqn{fluc}, reads
\be {\cal D}_4^2 \Gamma-[\phi
,[\phi,\Gamma]]=e^2\sum_{i=1}^2\left(\{Q_iQ_i^\dagger,\Gamma\}
+2Q_i\Omega_i Q_i^\dagger\right)\ .\label{gf}\ee
Since a rotation of the form $V_1=V_2$ can always be undone by a
gauge transformation, we may choose to work in a gauge in which
$V_1=V_2^\dagger$, so that $\Omega_1=-\Omega_2\equiv\Omega$ and
$g\approx 1-2i\Omega$. Then the above equation has the simple
solution $\Gamma(x^4)=-(\phi(x^4)/m)\Omega$, and the zero modes
read,
\be \delta\phi=0\ \ ,\ \ \delta Q_1=i\frac{q_1}{m}(\phi-m)\Omega \
\ ,\ \ \delta Q_2=i\frac{q_2}{m}(\phi+m)\Omega \ \ \ {\rm and}\ \
\delta A_4=-\frac{\partial_4\phi}{m}\Omega\ . \label{fluk}\ee
The coefficient in front of the two-derivative terms of the $U(2)$
chiral Lagrangian is determined by the overlap of these zero
modes, yielding
\be {\cal
L}_{2}=-\frac{v^2}{4m}\,\Tr\,(g^{-1}\partial_\alpha
g)(g^{-1}\partial^\alpha g)\label{2deriv}\ .\ee 
This result for the low-energy dynamics of the domain wall was
previously derived in \cite{sy}.

For solitons with worldvolume dimension greater than one, there is
an extra term in the moduli space approximation that is usually
ignored since it is subleading in the derivative expansion. It
comes from the  field strength $F_{\alpha\beta}$, with $\alpha$
and $\beta$ indices along the soliton worldvolume. To see this,
we first need the equation of motion for the gauge fields
$A_\alpha$
\be {\cal D}_\mu F_{\alpha\mu}-i[\phi,{\cal D}_\alpha\phi]=ie^2 (
Q_i\,{\cal D}_\alpha Q_i^\dagger-({\cal D}_\alpha
Q_i)Q_i^\dagger)\ .\ee
This equation is solved by $A_\alpha = -[1+(\phi/m)]\,
\partial_\alpha\Omega$, where the constant piece has been
chosen so  that $A_\alpha$ is pure gauge at $x^4=\pm\infty$. With
this ansatz, we have ${\cal D}_\alpha Q_i=\delta
Q_i\,\partial_\alpha\Omega$, with the zero mode given by
\eqn{fluk} and one can check that the equation
above reduces to that of \eqn{gf}. While this form of $A_\alpha$
holds for small fluctuations about $g=1$, the generalization to
arbitrary $g$ is simply \cite{sy}
\be A_\alpha=\frac{i}{2}\left[1+\frac{\phi(x^4)}{m}\right]\
g^{-1}\partial_\alpha g \ee
which, indeed, becomes pure gauge as $x^4\rightarrow -\infty$ and
vanishes as $x^4\rightarrow +\infty$, ensuring that any field
strength is localized on the wall.
%
To see how it leads to a
four-derivative term, we compute the field strength\footnote{A
similar discussion holds for other solitons in gauge theories ---
for example, vortices, monopoles or instantons
--- when their worldvolume dimension is greater than one. In
general, if a soliton has collective coordinates $X^a$ then the
zero modes are a combination of variations with respect to $X^a$
together with an infinitesimal gauge transformation, e.g.
$\delta_aA_\mu=\partial A_\mu /\partial X^a +{\cal
D}_\mu\epsilon_a$ where $A_\mu$ are the gauge fields transverse to
the soliton worldvolume. The equations of motion for the gauge
fields $A_\alpha$ parallel to the soliton worldvolume are solved
to leading order by $A_\alpha=\epsilon_a\,\partial_\alpha X^a$,
resulting in the field strength
\be F_{\alpha\beta}=\left(\partial\epsilon_a/\partial
X^b-\partial\epsilon_b/\partial
X^a-i[\epsilon_a,\epsilon_b]\right)\,\partial_\alpha
X^a\partial_\beta X^b\ .\nn\ee
The curvature term in brackets is part of the "universal bundle"
defined in \cite{as}. Some properties of this object are detailed
in \cite{jerome}. Inserting this expression into the action will
lead to four-derivative terms.}
\be F_{\alpha\beta}=\frac{i}{4}\left(1-\frac{\phi^2}{m^2}\right)
[g^{-1}\partial_\alpha g,g^{-1}\partial_\beta g]\ .\ee
Inserting this into the original five-dimensional Lagrangian and
integrating over $x^4$ gives rise to the promised four-derivative
term,
\be {\cal L}_{4-\partial}= c\  \Tr\, [(g^{-1}\partial_\alpha
g),\,(g^{-1}\partial_\beta g)]^2\label{4deriv}\ee
where the coefficient $c$ is given by the integral,
\be
c=\frac{1}{64e^2}\int_{-\infty}^{+\infty}\left(1-\frac{\phi(x^4)^2}{m^2}\right)^2\
dx^4\ . \ee
Parametrically, we see that $c\sim L_{\rm wall}/e^2$.  The
numerical prefactor can be computed in the limit $e^2v^2\ll m^2$
where the domain wall has the three layer structure shown in
figure 1. Up to corrections of order $ev/m$, arising from the two
outer layers, the profile of $\phi$ can be taken to be piecewise
linear,
\be \phi(x^4)\approx\left\{\begin{array}{lc}\ \ \ \  m\ \ \ \ \ \
& x^4 < -\frac{m}{e^2v^2} \\ -e^2v^2\,x^4\ \ \ \ \ \  &
-\frac{m}{e^2v^2} < x^4 < \frac{m}{e^2v^2} \\ \ \ -m & x^4 >
\frac{m}{e^2v^2}
\end{array}\right. \ee
which gives us
\be c\approx \frac{1}{60}\,\frac{m}{e^4v^2}\ .\ee
Note that terms with four derivatives and higher will also
typically appear if we go beyond the moduli space approximation,
taking into account the backreaction of the motion
$\partial_\alpha X^a$ on the fields themselves. The computation of
these is a much harder problem that we do not attempt here.

\subsection*{\it Domain Wall Skyrmions}

One can see at a glance from  the figure that, inside the domain
wall, the $U(2)$ gauge symmetry is restored. One may suspect that
this allows us to place an instanton, which in five-dimensions is
a particle-like soliton, inside the domain wall as long as its
size is smaller than $L_{\rm wall}\sim 2m/e^2v^2$. In
a supersymmetric theory, the instanton embedded within the domain
wall is not BPS \cite{18}. A related fact is that such a
configuration is not necessarily a stable solution because the
long range fields of the instanton are subject to the boundary
conditions imposed by the bulk. Since the vacuum is in the Higgs
phase, these boundary conditions are those of a five-dimensional
superconductor, ensuring that the magnetic field $F_{\mu\nu}$ with
$\mu,\nu=1,2,3,4$ must lie parallel to the surface of the wall:
$\partial_4F_{\mu\nu}$ at the boundary. We can ask how such a
configuration is viewed from the domain wall theory. 

Far from the core of the instanton, up to a power-law tail, the
gauge field configuration is pure gauge. We may write $A_\mu=-i
(\partial_\mu\,U)U^\dagger$, $\mu=1,2,3,4$. The instanton number $I$ is given
by the integral
\be I
= \frac{1}{24\pi^2}\,\int_{{\bf R}^3_{R}-{\bf
R}^3_{L}}d^3x\,\Tr\,\left[(\partial_\nu U)U^{-1}(\partial_\rho U)U^{-1}
(\partial_\sigma U)U^{-1}\right]\epsilon^{4\nu\rho\sigma} \ee
where we evaluate the integral over the two left and right
boundaries ${\bf R}^3_{L/R}$  of the domain wall. (These actually
have width equal to the penetration depth $1/ev$). But we've seen
above that such a large gauge transformation must be compensated
by a flavor transformation outside the domain wall in order to
leave the vacuum invariant. Thus $U\rightarrow V_1$ on ${\bf
R}^3_L$ and $U\rightarrow V_2$ on ${\bf R}^3_R$. We may work in
gauge in which $V_1=1$, while $V_2$ varies, in which case the
integral above does not pick up a contribution from ${\bf R}^3_L$
while, on ${\bf R}^3_R$, $U=V_2=g$, giving us
\be I=\frac{1}{24\pi^2}\,\Tr\,\int d^3x \,\left[(\partial_\alpha
g)g^{-1}(\partial_\beta g)g^{-1}(\partial_\gamma
g)g^{-1}\right]\epsilon^{\alpha\beta\gamma}=B\ee
where $\alpha,\beta,\gamma$ run over $1,2,3$, the spatial indices
of the domain wall. The above expression measures the baryon
number $B$ of the Skyrme model, reiterating the result of Atiyah
and Manton \cite{am}: a Yang-Mills configuration of instanton
charge $I$ maps into a Skymre configuration with baryon number
$B=I$.

So much for topological charges. The real surprise of Atiyah and
Manton is that the instanton solution gives such a good
approximation to the Skyrmion. How can we understand this from our
set-up? The Skyrmion solution balances the two-derivative
\eqn{2deriv} and four derivative \eqn{4deriv} terms against each
other, so that they are comparable. This means that the fields of
the Skyrmion vary with wavelength,
\be L_{\rm Skyrmion}\sim(\partial g)^{-1} \sim m/e^2v^2\sim L_{\rm
wall} \ee
suggesting that an instanton much smaller than $L_{\rm wall}$ will
expand until it just touches the wall. Indeed, Atiyah and Manton
showed the Skyrme functional is minimized by an instanton of a
particular size which, from our perspective, is the width of the
wall $L_{\rm wall}$.

Let's look at this in more detail. There is a Bogomolnyi type argument bounding
the mass of the Skyrmion: $M_{\rm Skyrme}\geq M_{\rm bound}$.
However, unlike BPS solitons, the bound is never saturated.
Numerical study reveals that for a single $B=1$ Skyrmion, $M_{\rm
Skyrme}\sim 1.2\times M_{\rm bound}$. Using the scaling above, and
working in the $e^2v^2\ll m^2$ limit, the bound on the Skyrmion
mass is
\be M_{\rm bound}\approx
48\pi^2\sqrt{2\,\frac{v^2}{4m}\,\frac{m}{60e^4v^2}}\ B\approx
\frac{4.4\pi^2}{e^2}\,I\approx 1.1\times M_{\rm instanton}\ .\ee
where, in our conventions, the mass of a single instanton is
$M_{\rm inst}=4\pi^2/e^2$. We see that the true mass of the
Skyrmion in the domain wall is about $30\%$ higher than that of
the instanton in Coulomb phase. Nevertheless, the fact that the
Atiyah-Manton construction \eqn{am} results in a good
approximation to the Skymion at the $1\%$ level suggests that the
instanton solution is not greatly deformed by its residence inside
the domain wall. It would be interesting to use this perspective
to quantify the Atiyah-Manton approximation.

Note that the above analysis is from the perspective of the
domain wall and its validity becomes dubious
for instantons much smaller than $L_{\rm wall}$ since, for
$(\partial g)^{-1} \sim L_{\rm wall}$, one might worry about
higher derivative terms in the effective Lagrangian.
This is the standard drawback of the Skyrme model: if the two and
four derivative terms are of equal parametric importance, then so
are all higher derivatives. In the present context, our small
dimensionless parameter $e^2v^2/m^2$ may mitigate matters. Dimensional
analysis requires higher derivative terms to be of the form 
$b_nmv^2 (\partial/m)^{2n}$, where $b_n$ is a dimensionless
coefficient. However, to compete with the two and
four derivative terms, one would need $b_n\sim(m^2/e^2v^2)^{2n-2}$ and
it is not clear that such large coefficients arise. Further study of this
issue would be worthwhile.

\subsection*{\it Generalizations}

There exists an obvious generalization of our analysis to higher
gauge groups $SU(N)$. One starts with a five dimensional $U(N)$
gauge theory with $2N$ flavors. As above, the first $N$ flavors
are assigned mass $(\phi-m)$, with the remaining assigned mass
$(\phi+m)$. There exists a domain wall interpolating between the
two vacua $\phi=\pm m{\bf 1}_N$, whose low-energy dynamics is
described by the $SU(N)$ Skyrme model, complete with four
derivative term. As above, the instantons of the gauge theory
descend to Skyrmions on the domain wall.

For $SU(N)$ gauge theories with $N\geq 3$, there exists a further
term that one can add to the five dimensional action: a
Chern-Simons coupling, with gauge invariance enforcing an
integer valued coefficient $k$ \cite{ims}.
%
%
How does such a
term appear in the domain wall worldvolume theory? The answer is
that induces a Wess-Zumino term for the Skyrme model. As is well
known, a Lagrangian for such a term usually involves introducing a
five-manifold $M$ with boundary $\partial M\cong {\bf R}^{1,3}$
\cite{witten}.
%
%
From our perspective, the five dimensional manifold appears
naturally as the bulk, with the domain wall acting as the
boundary. Such a term determines the spin properties of the
Skyrmion: fermion for $k$ odd, boson for $k$ even. It should be
possible to repeat this soliton quantization from the perspective of the
instanton in five dimensions.

Finally, let us mention that a Chern-Simons term is also induced
in five dimensional gauge theories by integrating out massive
Dirac fermions. If these also couple to the scalar field $\phi$,
then two effects occur: firstly, there are chiral, fermionic zero
modes localized on the domain wall. Secondly, they induce a term
of the form $\phi F_{\mu\nu}F^{\mu\nu}$, which is related to the
Chern-Simons term by supersymmetry. It would be interesting to understand
the interplay of these two effects in the dynamics of the domain
wall.

\newpage

\section*{Acknowledgement}
We would like to thank Sean Hartnoll, Matt Headrick, Youichi Isozumi, Nick
Manton and Norisuke Sakai for useful discussions. The work of M.~N. and K.~O. (M.~E.)
is supported by Japan Society for the Promotion of Science under
the Post-doctoral (Pre-doctoral) Research Program. D.~T. is
supported by the Royal Society and thanks the Benasque and Aspen
centers for physics for hospitality while this work was completed.

\end{document}